\documentclass[12pt]{article}
\usepackage{graphicx}
\usepackage{epstopdf}
\usepackage{amsmath}
\usepackage{amsfonts}
\usepackage{amssymb}
\usepackage{color}
\usepackage{mathrsfs}
\usepackage{multirow}

\newcommand{\be}{\begin{equation}}
\newcommand{\ee}{\end{equation}}
\newcommand{\barr}{\begin{array}}
\newcommand{\earr}{\end{array}}

\usepackage{latexsym}
\usepackage{amsthm}
\usepackage{amsmath}
\usepackage{graphicx}
\usepackage{amssymb}
\usepackage{bbm}
\newcommand{\gsim}{\lower.7ex\hbox{$\;\stackrel{\textstyle>}{\sim}\;$}}
\newcommand{\lsim}{\lower.7ex\hbox{$\;\stackrel{\textstyle<}{\sim}\;$}}

\newcommand{\bea}{\begin{eqnarray}}
\newcommand{\eea}{\end{eqnarray}}

\newcommand{\comment}[1]{}

\def\e{\mathrm{e}}

\def\half{{1\over 2}}

\def\half{{1\over 2}}
\def\({\left(}
\def\){\right)}
\def\[{\left[}
\def\]{\right]}
\def\e{\begin{equation}}
\def\q{\end{equation}}
\def\m{\begin{eqnarray}}
\def\n{\end{eqnarray}}

\begin{document}


\setcounter{page}{1} \baselineskip=15.5pt \thispagestyle{empty}

\begin{flushright}
\end{flushright}
\vfil

\begin{center}

{\Large \bf Constraint on the primordial gravitational waves \\
\vspace{1mm}
from the joint analysis of BICEP2 \\
\vspace{3mm}
and Planck HFI 353 GHz dust polarization data}
\\[0.7cm]
{Cheng Cheng $^{\Diamond,\ \heartsuit}$ \footnote{chcheng@itp.ac.cn}, Qing-Guo Huang $^\Diamond$ \footnote{huangqg@itp.ac.cn} and Sai Wang $^\Diamond$ \footnote{wangsai@itp.ac.cn}}
\\[0.7cm]

{\normalsize { \sl $^\Diamond$ State Key Laboratory of Theoretical Physics, Institute of Theoretical Physics, \\ Chinese Academy of Science, Beijing 100190, China}}\\
\vspace{.2cm}

{\normalsize { \sl $^\heartsuit$  University of the Chinese Academy of Sciences, Beijing 100190, China}}
\vspace{.3cm}

\end{center}

\vspace{.8cm}

\hrule \vspace{0.3cm}
{\small  \noindent \textbf{Abstract} \\[0.3cm]
\noindent We make a joint analysis of BICEP2 and recently released Planck HFI 353 GHz dust polarization data, and find that there is no evidence for the primordial gravitational waves and the bound on the tensor-to-scalar ratio becomes $r<0.083$ at $95\%$ confidence level in the base $\Lambda$CDM + tensor model. Extending to the model with running of scalar spectral index, the bound is a little bit relaxed to $r<0.116$ at $95\%$ confidence level. Our results imply that the inflation model with a single monomial potential is marginally disfavored at around $95\%$ confidence level. Especially, the $m^2\phi^2/2$ inflation model is disfavored at more than $2\sigma$ level. However, the Starobinsky inflation model gives a nice fit.
}
 \vspace{0.3cm}
\hrule

\begin{flushleft}
\end{flushleft}

\vspace{8cm}

\newpage
\section{Introduction}

In the last decades, inflation \cite{Guth:1980zm,Linde:1981mu,Albrecht:1982wi} was taken as the leading paradigm for the very early universe. Not only does it solve the flatness, horizon and monopole problems, it also provides the seeds for the anisotropies in the cosmic microwave background (CMB) and the formation of large-scale structure. The simplest inflation models predict that the spectra of both the scalar perturbations and primordial gravitational waves are adiabatic, Gaussian and nearly scale-invariant.
The primordial gravitational waves can contribute to the B-mode polarization pattern.
Thus the detection of an excess of B-mode power over the base lensed-$\Lambda$CDM expectation is taken as the direct detection of primordial gravitational waves \cite{Grishchuk:1974ny,Starobinsky:1979ty,Rubakov:1982,Crittenden:1993ni,Kamionkowski:1996zd,Kamionkowski:1996ks,Hu:1997mn}.

In this March, BICEP2 collaboration released its data and claimed that an excess of B mode power over the base lensed-$\Lambda$CDM expectation in the range $30< \ell <150$ was detected at a significance of $>5\sigma$ \cite{Ade:2014xna}. BICEP2 collaboration also claimed that the observed B mode power spectrum is well fitted by a lensed-$\Lambda$CDM+tensor cosmology with the tensor-to-scalar ratio $r=0.2_{-0.05}^{+0.07}$ and $r=0$ disfavored at $7\sigma$. However, there were debates on whether this signal could be interpreted as primordial gravitational waves in the last few months. Denote $D_\ell\equiv \ell(\ell+1)C_\ell/(2\pi)$.
In \cite{Mortonson:2014bja}, based on the model of $D_\ell^{\rm BB,dust}\propto \ell^{-0.42}$, Mortonson and Seljak found that the joint BICEP2+Planck analysis prefers that the B mode detected by BICEP2 should be explained as the polarized dust, not the primordial gravitational waves, and the upper limit of the tensor-to-scalar ratio is $r<0.11$ at $95\%$ C.L..
Roughly at the same time, in \cite{Flauger:2014qra}, Flauger, Hill and Spergel also pointed out that the signal of BICEP2 can be interpreted as dust without primordial gravitational waves. Furthermore, they gave four independent estimates of the dust polarization in the observed BICEP2 field, and concluded that BICEP2 data alone cannot distinguish between forgrounds and a primordial gravitational wave signal.
Mapping the B modes in the BICEP2 region from the publicly available Q and U 353 GHz preliminary Planck polarization maps, Colley and Gott found a positive correlation coefficient of $(15.2\pm 3.9)\%$ $(1\sigma)$ which implies that a gravitational wave signal with a root-mean-square amplitude equals to $54\%$ of the total BICEP2 root-mean-square amplitude, or equivalently $r=0.11\pm 0.04$, in \cite{Colley:2014nna}.

Recently Planck collaboration released the Planck HFI polarization data from 100 to 353 GHz to measure the polarized dust angular power spectra $C_\ell^{\rm EE}$ and $C_\ell^{\rm BB}$ over the multipole range $40<\ell<600$ well away from the Galactic plane in \cite{Adam:2014oea}. Extrapolating the Planck 353 GHz data to 150 GHz, Planck collaboration gave a dust power $D_\ell^{\rm BB,dust}=1.32\times 10^{-2}$ $\mu$K$^2$ over the multipole range $40<\ell<120$ with the statistical uncertainty $\pm 0.29\times 10^{-2}$ $\mu$K$^2$ and an additional uncertainty $(+0.28,-0.24)\times 10^{-2}$ $\mu$K$^2$ from the extrapolation. The dust power is roughly the same magnitude as BICEP2 signal!

In this paper we will make a joint analysis of BICEP2 and recently released Planck HFI 353 GHz dust polarization data to work out the constraint on the tensor-to-scalar ratio. Our paper will be organized as follows. In Sec.~2, we will explain the methodology we adopt and the results of our global fitting. We use our fitting results to constrain several typical inflation models in Sec.~3. Discussion will be given in Sec.~4.

\section{Data analysis}

\subsection{Methodology}

It is convenient to parametrize the power spectra of scalar and tensor perturbations as follows
\m
P_s(k)&=&A_s \({k\over k_p}\)^{n_s-1+\half {dn_s\over d\ln k}\ln{k\over k_p}},\\
P_t(k)&=&A_t \({k\over k_p}\)^{n_t},
\n
where $A_s$ and $A_t$ are the amplitudes of scalar and tensor power spectra at the pivot scale $k_p$, $n_s$ and $n_t$ are the spectral indices of scalar and tensor power spectra, and $dn_s/d\ln k$ is the running of scalar spectral index. Here the pivot scale is set to be $k_p=0.05$ Mpc$^{-1}$, and the tilt of tensor power spectrum is related to $A_t$ and $A_s$ by  $n_t=-A_t/(8A_s)$ \footnote{One may take $n_t$ as a free parameter. But, in fact, $n_t$ cannot be significantly constrained by the current data at all. }.

We add a prior for counting the recently released Planck HFI 353 GHz dust polarization data into the Markov Chain Monte Carlo sampler, namely CosmoMC \cite{Lewis:2002ah}, to explore the cosmological parameters space. The combined datasets with BICEP2 BB \cite{Ade:2014xna}, Planck TT \cite{Ade:2013zuv}, WMAP Polarization \cite{Hinshaw:2012aka} and Planck 353 GHz HFI polarization data \cite{Adam:2014oea} will be used in our global fitting.

In this section we consider two cosmological models. \\
$\bullet$ One is the base $\Lambda$CDM+tensor cosmology where there are seven parameters: the baryon density today $(\Omega_b h^2)$, the cold dark matter density today $(\Omega_c h^2)$, the $100\times$ angular scale of the sound horizon at last-scattering ($100\theta_{\rm MC}$), the Thomson scattering optical depth due to the reionization $(\tau)$, the amplitude of scalar power spectrum $(A_s)$, the spectral index of scalar power spectrum $(n_s)$ and the tensor-to-scalar ratio $(r)$. \\
$\bullet$ The other is the $\Lambda$CDM+nrun+tensor model where one extra parameter, namely the running of scalar spectral index $(dn_s/d\ln k)$, is added.


\subsection{Global fitting}

Our results are accumulated in Table \ref{table:fitting}.
\begin{table*}[!hts]
\footnotesize
\centering
\renewcommand{\arraystretch}{1.5}
\begin{tabular}{c|c|c|c|c}
\hline \hline
\multirow{2}{2cm}{parameter}& \multicolumn{2}{c|} {$\Lambda$CDM+tensor}&\multicolumn{2}{c}{$\Lambda$CDM+nrun+tensor} \\
\cline{2-5}
 &+$\sigma_{\rm stat}$ &+$\sigma_{\rm stat+extr}$ &+$\sigma_{\rm stat}$ &+$\sigma_{\rm stat+extr}$\\
\hline
$\Omega_b h ^2$ &$0.02193\pm 0.00027$&$0.02194\pm 0.00027$ &$0.02212\pm 0.00030$ &$0.02213\pm 0.00030$  \\
$\Omega_c h ^2$ &$0.1198_{-0.0025}^{+0.0026}$ &$0.1197\pm 0.0026$ &$0.1205\pm 0.0027$ &$0.1205^{+0.0027}_{-0.0026}$ \\
$100\theta_{MC}$ &$1.04107_{-0.00064}^{+0.00062}$ &$1.04110\pm0.00062$  &$1.04109 \pm0.00063$ &$1.04111\pm 0.00063$ \\
$\tau$ &$0.089^{+0.013}_{-0.014}$ &$0.090^{+0.012}_{-0.014}$ &$0.099^{+0.014}_{-0.016}$ &$0.099^{+0.014}_{-0.017}$ \\
$\ln(10^{10}A_s)$ &$3.086^{+0.024}_{-0.027}$ &$3.087^{+0.024}_{-0.026}$ &$3.111^{+0.027}_{-0.033}$ &$3.113^{+0.029}_{-0.034}$ \\
$n_s$ &$0.9577\pm 0.0071$ &$0.9580_{-0.0071}^{+0.0072}$ &$0.9528^{+0.0078}_{-0.0077}$ &$0.9528\pm0.0078$ \\
$d n_s/d \ln k$ & - & - &$-0.018^{+0.010}_{-0.009}$ &$-0.018\pm 0.009$ \\
$r_{0.05}$ ($95\%$ C.L.) &$<0.083$  &$<0.083$ &$<0.117$ &$<0.116$ \\
\hline
\end{tabular}
\caption{The $68\%$ limits for the cosmological parameters in the base $\Lambda$CDM+tensor and $\Lambda$CDM+nrun+tensor models. }
\label{table:fitting}
\end{table*}

\begin{figure}[!htb]
\begin{center}
\includegraphics[width=12 cm]{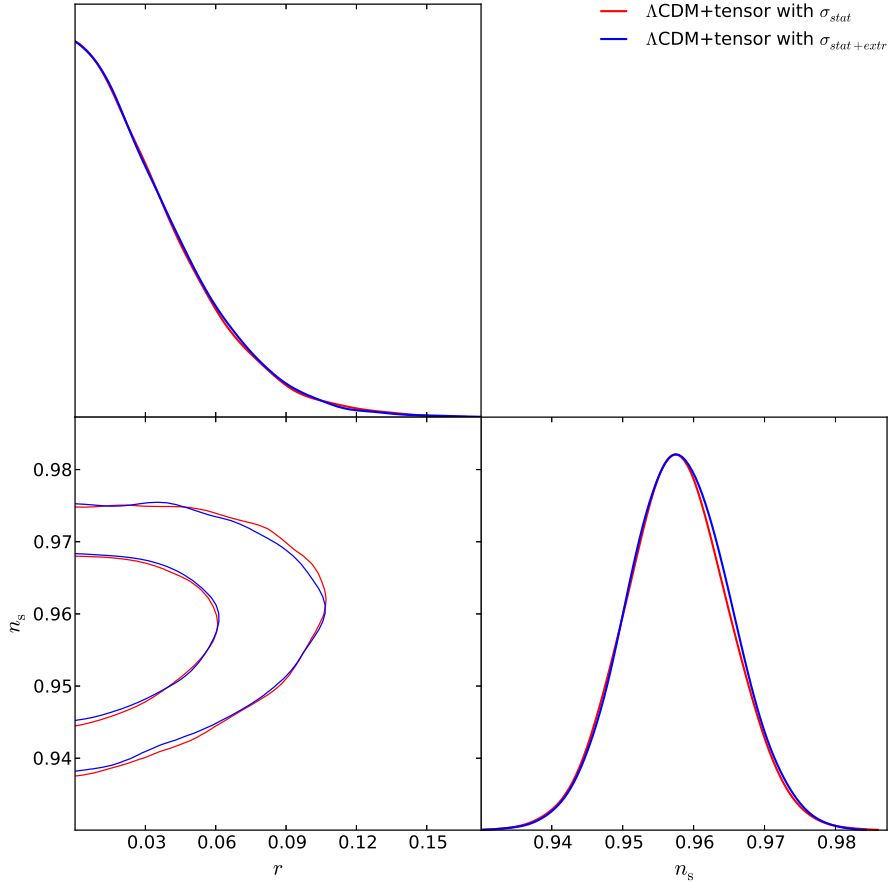}
\caption{The marginalized contour plot and the likelihood distributions of $r$ and $n_s$ in the base $\Lambda$CDM+tensor cosmology. }
\end{center}
\label{fig:lcdm}
\end{figure}

The marginalized contour plot and the likelihood distributions of $r$ and $n_s$ in the base $\Lambda$CDM+tensor cosmology are illustrated in Fig.~\ref{fig:lcdm}.
We see that the constraints on the tensor-to-scalar ratio $r$ and the scalar spectral index $n_s$ are $r_{0.05}<0.083$ ($95\%$ C.L.)
and $n_s=0.9577\pm 0.0071$ ($68\%$ C.L.) if only the statistical uncertainty in dust power are considered,
and $r_{0.05}<0.083$ ($95\%$ C.L.) and $n_s=0.9580^{+0.0072}_{-0.0071}$ ($68\%$ C.L.)
if the additional uncertainty is taken into account from the spectral extrapolation.
The primordial scalar power spectrum deviates from the Harrison-Zel'dovich spectrum at around $6\sigma$ level.

\begin{figure}[!htb]
\begin{center}
\includegraphics[width=16 cm]{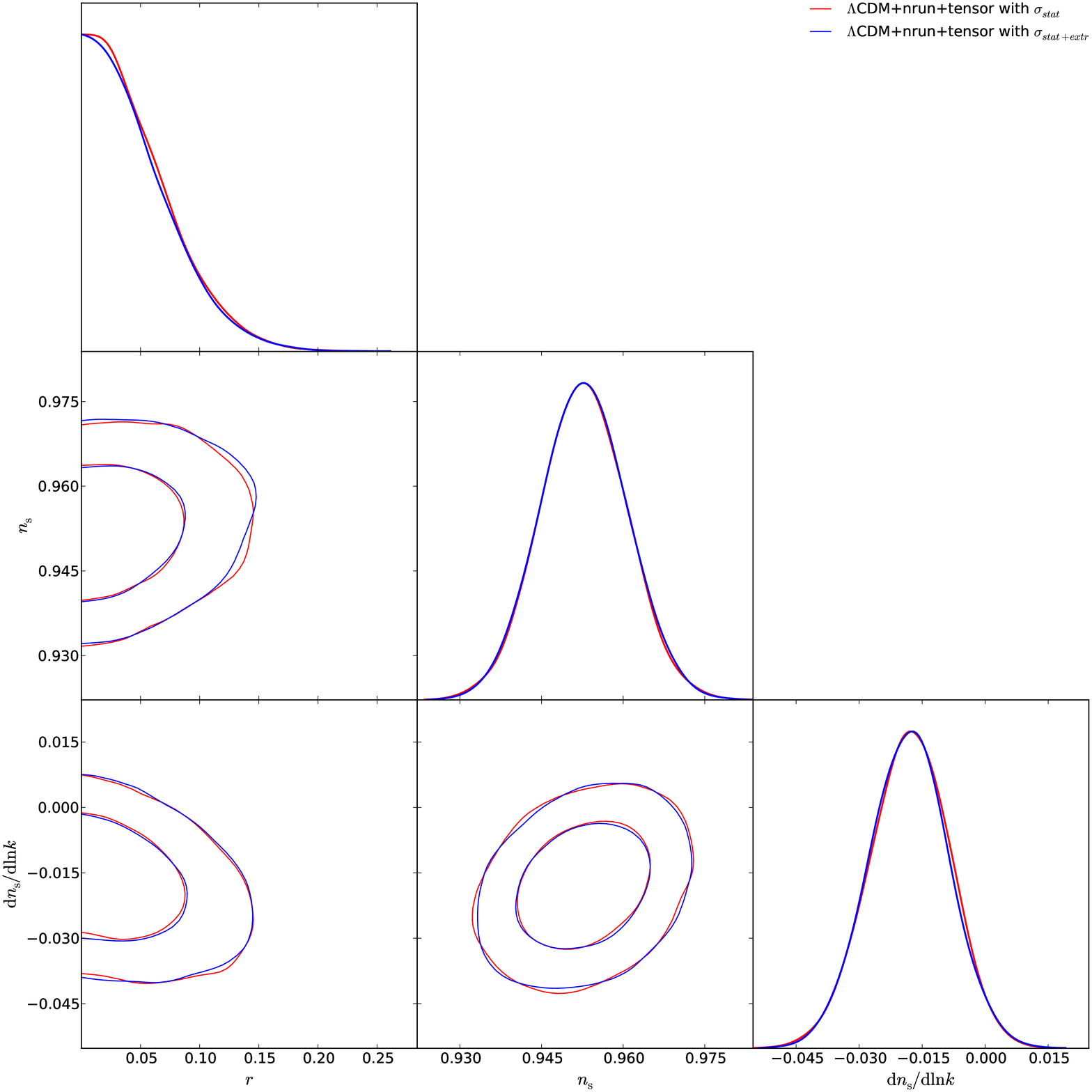}
\end{center}
\caption{The marginalized contour plot and the likelihood distributions of $r$, $n_s$ and $dn_s/d\ln k$ in the base $\Lambda$CDM+nrun+tensor cosmology. }
\label{fig:lcdmnrun}
\end{figure}

Similarly, the marginalized contour plots and the likelihood distributions of $r$, $n_s$ and $dn_s/d\ln k$ in the base $\Lambda$CDM+nrun+tensor cosmology are illustrated in Fig.~\ref{fig:lcdmnrun}.
We see that the constraints on the tensor-to-scalar ratio $r$, the scalar spectral index $n_s$ and its running $dn_s/d\ln k$
are $r_{0.05}<0.117$ ($95\%$ C.L.), $n_s=0.9528^{+0.0078}_{-0.0077}$ ($68\%$ C.L.)
and $dn_s/d\ln k=-0.018^{+0.010}_{-0.009}$ ($68\%$ C.L.) if only the statistical uncertainty
in dust power are considered, and $r_{0.05}<0.116$ ($95\%$ C.L.), $n_s=0.9528\pm 0.0078$ ($68\%$ C.L.)
and $dn_s/d\ln k=-0.018\pm 0.009$ ($68\%$ C.L.) if the additional uncertainty is taken into account from the spectral extrapolation.
By contrast to the base $\Lambda$CDM+tensor model, $\Delta \chi^2 = -2.2$ and $-0.6$ respectively in the above two fits.
The observational data marginally prefers a negative running of scalar spectral index at around $2\sigma$ level.
The primordial scalar power spectrum at the pivot scale $k_p=0.05$ Mpc$^{-1}$ deviates from the Harrison-Zel'dovich spectrum at around $6\sigma$ level as well.

\section{Constraint on inflation models}

The simplest inflation model is the so-called canonical single-field slow-roll inflation model which is governed by a dynamical canonical scalar field $\phi$. The equations of motion of the homogeneous background during inflation are given by
\m
\ddot \phi+3H\dot \phi+V'(\phi)=0,
\n
and
\m
H^2={1\over 3M_p^2}\(\half \dot \phi^2+V(\phi)\),
\n
where $M_p=(8\pi G)^{-\half}$ is the reduced Planck scale, $V(\phi)$ is the potential of inflaton field $\phi$ and the prime denotes the derivative with respective to $\phi$. The inflaton field slowly rolls down its potential if $\epsilon\ll1$ and $|\eta|\ll 1$, where
\m
\epsilon\equiv {M_p^2\over 2}\({V'\over V}\)^2,\quad \eta\equiv M_p^2 {V''\over V}.
\n
The parameters of the scalar and tensor power spectra are evaluated at the value of $\phi$ when the perturbation mode $k$ exits the horizon during inflation, namely $k=aH$. The e-folding number before the end of inflation at which the pivot scale $k_p$ crosses the horizon is roughly related to the value of $\phi$ by
\m
N\simeq {1\over M_p^2}\int^{\phi_N}_{\phi_{\rm end}} {V\over V'}d\phi.
\n
Usually the e-folding number before the end of inflation corresponding to the pivot scale is taken between 50 and 60 \cite{Liddle:2003as}. The amplitudes of the scalar and tensor power spectra are given by
\m
P_s&=& {V\over 24\pi^2 M_p^4 \epsilon}, \\
P_t&=& {2V\over 3\pi^2 M_p^4}.
\n
Therefore the tensor-to-scalar ratio, the spectral index and its running of scalar power spectral index and the spectral index of tensor power spectrum are
\m
r&=&16\epsilon, \\
n_s&=&1-6\epsilon+2\eta, \\
{dn_s\over d\ln k}&=&-24\epsilon^2+16\epsilon\eta-2\xi, \\
n_t&=& -2\epsilon,
\n
where
\m
\xi\equiv M_p^4{V'V'''\over V^2}.
\n
See the appendix of \cite{Huang:2014yaa} for more accurate formula. From the above formula, there is a consistency relation for the canonical single-field slow-roll inflation model, i.e.
\m
n_t=-r/8.
\n

\begin{figure}[!htb]
\begin{center}
\includegraphics[width=16 cm]{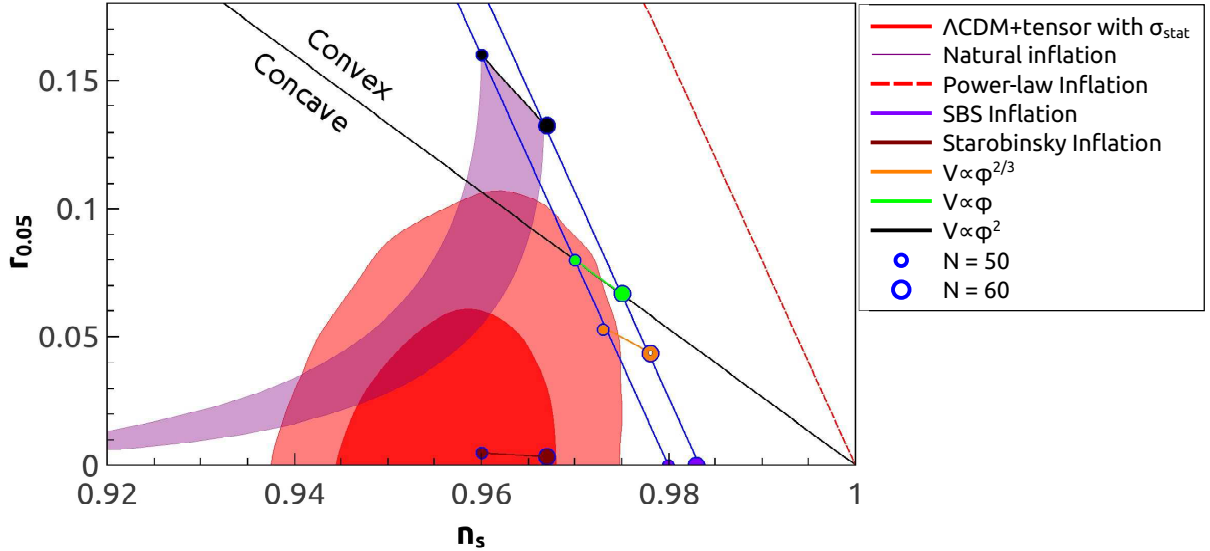}
\caption{Comparing the typical inflation models with the observational constraints. }
\end{center}
\label{fig:comparison}
\end{figure}

Here we will consider several typical inflation models and compare them to the constraints obtained in the previous section. See Fig.~\ref{fig:comparison}.
From this figure, the inflation model with a concave potential is preferred at around $95\%$ confidence level.

Power law potential and chaotic inflation model \cite{Linde:1983gd}. The inflation model with power law potential $V(\phi)\sim \phi^n$ is the simplest class of inflation models, and is the prototype of chaotic inflation model. \footnote{Thank A.~Linde for clarifying this point to us. See more discussion about chaotic inflation in \cite{Kallosh:2014xwa}.}
In string theory, a general mechanism for chaotic inflation was proposed to be driven by monodromy axion field. For example, $n=2/3,\ 2/5$ in \cite{Silverstein:2008sg}, $n=1$ in \cite{McAllister:2008hb}, and higher power can be obtained in \cite{Marchesano:2014mla,McAllister:2014mpa}. In general, the model with potential $V(\phi)\sim \phi^n$ predicts
\m
r&=&{4n\over N}, \\
n_s&=&1-{n+2\over 2N}.
\n
From Fig.~\ref{fig:comparison}, we see that $m^2\phi^2/2$ chaotic inflation model is disfavored at more than $2\sigma$ level, and almost all of the inflation models with a single monomial potential are marginally disfavored at around $2\sigma$ level.

Natural inflation model \cite{Freese:1990rb,Adams:1992bn}. In this model, the effective one-dimensional potential is given by
\m
V(\phi)=m^2f^2 \(1+\cos {\phi\over f}\),
\n
where $f$ is the so-called decay constant.
This inflation model predicts
\m
n_s&=&1-{1\over (f/M_p)^2}{3+\cos\theta_N\over 1-\cos\theta_N}, \\
r&=&{8\over (f/M_p)^2}{1+\cos \theta_N\over 1-\cos \theta_N},
\n
where
\m
\cos{\theta_N\over 2}=\exp \(-{N\over 2(f/M_p)^2}\).
\n
In the limit of $f/M_p\gg 1$, $\theta_N\simeq \sqrt{4N}M_p/f$, and hence $r=8/N$ and $n_s=1-2/N$ which are the same as those in the chaotic inflation model with potential $V(\phi)=\half m^2\phi^2$.
Compared to the observational data, Natural inflation is still compatible with the data. See shaded purple region in Fig.~\ref{fig:comparison}.

Power-law inflation model \cite{Lucchin:1984yf}. The inflaton potential takes the form
\m
V(\phi)=V_0\exp \(-\sqrt{2\over p}{\phi\over M_p}\).
\n
The tensor-to-scalar ratio and the scalar spectral index in the power-law inflation model are given by
\m
r&=&{16\over p}, \\
n_s&=& 1-{2\over p},
\n
and then $r=8(1-n_s)$. The prediction of power-law inflation model is illustrated as the red dashed line in Fig.~\ref{fig:comparison} which is disfavored at more than $2\sigma$ level.

Starobinsky inflation model \cite{Starobinsky:1980te}. In this model, the inflationary expansion of the universe is driven by the higher derivative term and its action is
\m
S={1\over 16\pi G}\int d^4 x\sqrt{-g} \(R+{R^2\over 6M^2}\),
\n
where $M$ is an energy scale. The predictions of Starobinsky inflation model are
\m
r&=&{12\over N^2}, \\
n_s&=&1-{2\over N},
\n
in \cite{Mukhanov:1981xt,Starobinsky:1983zz}. From the viewpoint of fundamental theory, the higher-power terms are also expected, and the predictions are given in \cite{Huang:2013hsb}. Fig.~\ref{fig:comparison} indicates that this model can fit the data quite well.

Spontaneously broken SUSY (SBS) inflation model \cite{SUSY}. The potential of inflaton field in SBS inflation model is given by
\m
V(\phi)=V_0\(1+c\ln {\phi\over Q}\),
\n
where $V_0$ is dominant and $c\ll 1$. This model preditcs
\m
r&\simeq &0, \\
n_s&=&1-{1\over N}.
\n
Since a large scalar spectral index is predicted in SBS inflation model, it is disfavored at more than $2\sigma$ level.

\section{Discussion}

In this paper we take into account the Planck HFI 353 GHz polarization data and work out the constraint on the tensor-to-scalar ratio by combining with BICEP2 BB, Planck TT and WMAP Polarization datasets. We do not find any evidence for the primordial gravitational waves, and the upper limits of the tensor-to-scalar ratio are given by $r<0.083$ $(95\%)$ C.L. in the base $\Lambda$CDM+tensor model, and $r<0.116$ $(95\%)$ C.L. in the base $\Lambda$CDM+nrun+tensor model. Our results are tighter than those from Planck TT+WMAP Polarization \cite{Ade:2013zuv}. But there is still a room for the detectable signal of gravity wave in the near future.

Comparing the inflation models with the observational constraints obtained in this paper, we find that the Starobinsky inflation model gives a nice fit,
but the inflation model with a single monomial potential is marginally disfavored at around $95\%$ C.L..
Especially, the $m^2\phi^2/2$ inflation model is disfavored at more than $2\sigma$ level,
as well as the SBS inflation model and the power-law inflation model.
The natural inflation model is still compatible with the data.
The inflation model with a concave potential is preferred at around $95\%$ C.L..

Given the uncertainties in the amplitude of Planck dust polarization, we cannot determine whether the B-mode polarization reported by BICEP2 stems from the primordial gravitational waves or the polarized dust. A cross-correlation between BICEP2 and Planck datasets might be able to reduce the uncertainties in the amplitude of Planck dust polarization and then improve the constraint on the tensor-to-scalar ratio.
Thus we believe that a careful joint analysis of BICEP2 and Planck data, in particular masking, filtering and color corrections, is still needed in the near future. We hope that it will be done soon.

\vspace{5mm}
\noindent{\large \bf Acknowledgments}

We acknowledge the use of Planck Legacy Archive, ITP and Lenovo
Shenteng 7000 supercomputer in the Supercomputing Center of CAS
for providing computing resources.
This work is supported by the project of Knowledge Innovation Program of Chinese Academy of Science and grants from NSFC (grant NO. 11322545 and 11335012).

\newpage

\end{document}